\documentclass{elsart}

\usepackage{HEP5} 
\usepackage{amsmath}
\usepackage{amssymb}
\usepackage{amsfonts}
\usepackage{amscd}
\usepackage{bbm} 
\usepackage{fleqn} 
\usepackage[footnotesize]{caption}
\usepackage{graphicx} 
\usepackage{braket} 
\usepackage{mathrsfs}
\usepackage[center,footnotesize,hang]{subfigure}
\usepackage[bbgreekl]{MyBbol}
\usepackage{longtable}
\usepackage{accents}
\allowdisplaybreaks[1] 
\newcommand{\CenterEps}[2][1]{
\ensuremath{\vcenter{\hbox{\includegraphics[scale=#1]{#2.eps}}}}
}

\newcommand{\RaiseBrace}[1]{\raise3pt\hbox{$\displaystyle#1$}}

\def\D{\mathrm{d}}

\def\<{\left\langle}
\def\>{\right\rangle}

\DeclareMathOperator{\diag}{diag}

\begin{document}

\begin{frontmatter}

\begin{flushright}
{\small TUM-HEP-466/02}
\end{flushright}
\vspace*{2.cm}
\title{
The LMA Solution from Bimaximal Lepton Mixing at the GUT Scale by 
Renormalization Group Running
}
\author{Stefan Antusch\thanksref{label01}},
\thanks[label01]{E-mail: \texttt{santusch@ph.tum.de}}
\author{J\"{o}rn Kersten\thanksref{label04}},
\thanks[label04]{E-mail: \texttt{jkersten@ph.tum.de}}
\author{Manfred Lindner\thanksref{label05}},
\thanks[label05]{E-mail: \texttt{lindner@ph.tum.de}}
\author{Michael Ratz\thanksref{label06}}
\thanks[label06]{E-mail: \texttt{mratz@ph.tum.de}}
\address{Physik-Department T30, 
Technische Universit\"{a}t M\"{u}nchen\\ 
James-Franck-Stra{\ss}e,
85748 Garching, Germany
}

\begin{abstract}
We show that in see-saw models with bimaximal lepton mixing at
the GUT scale and with zero CP phases, the
solar mixing angle $\theta_{12}$ generically evolves 
towards sizably smaller values due to Renormalization Group effects,
whereas the evolution of $\theta_{13}$ and $\theta_{23}$ is
comparatively small. 
The currently favored LMA solution of the solar neutrino problem 
can thus be obtained in a natural way from bimaximal mixing
at the GUT scale. 
We present numerical examples for the evolution of the leptonic mixing angles
in the Standard Model and the MSSM, 
in which the current best-fit values of the LMA mixing angles are produced.
These include a case where the mass eigenstates corresponding to the solar mass
squared difference have opposite CP parity.  
\end{abstract}

\end{frontmatter}

\newpage

\section{Introduction}
Recent experimental evidence strongly favors the LMA solution of the
solar neutrino problem with a large but non-maximal value of the solar
mixing angle $\theta_{12}$ 
\cite{Barger:2002iv,Bandyopadhyay:2002xj,Bahcall:2002hv,deHolanda:2002pp}.
An overview of the current allowed regions for the mixing angles 
and the mass squared differences is given in table \ref{tab:ExpData}. 
\begin{table}[h]
\begin{center}
\begin{longtable}{l|ccc}
& Best-fit value & Range (for $\theta \in [0^\circ,45^\circ]$) & C.L. \\
  \hline
$\theta_{12}$ [${\:}^\circ$]& $32.9$ & $26.1 - 43.3$ & $99 \% \;(3 \sigma)$\\  
$\theta_{23}$ [${\:}^\circ$]& $45.0$ & $33.2 - 45.0$ &    $99 \%\;(3 \sigma)$\\
$\theta_{13}$ [${\:}^\circ$]& $-$ &    $0.0-  9.2$ &     $90 \%\;(2 \sigma)$\\
$\Delta m^2_{\mathrm{sol}}$ [eV$^2$] & $5 \cdot 10^{-5}$&
$2.3\cdot 10^{-5} - 3.7\cdot 10^{-4}$&	$99 \%\;(3 \sigma)$ \\
$|\Delta m^2_{\mathrm{atm}}|$ [eV$^2$] & $2.5 \cdot 10^{-3}$&
$1.2\cdot 10^{-3} - 5\cdot 10^{-3}$ &	$99 \%\;(3 \sigma)$\hfill
\end{longtable}
\setcounter{table}{0}
\vspace{2mm}
\caption{
 Experimental data for the neutrino mixing angles and mass squared
 differences. 
 For the solar angle $\theta_{12}$ and the solar mass squared
 difference, the LMA solution has been assumed.
 The results stem from the analysis of the recent SNO data
 \cite{Bahcall:2002hv}, the Super-Kamiokande atmospheric data
 \cite{Toshito:2001dk} and the CHOOZ experiment \cite{Apollonio:1999ae}.
}
\label{tab:ExpData}
\end{center}
\end{table}

A big problem for model builders is to explain the
deviation of $\theta_{12}$ from maximal mixing, while keeping
$\theta_{23}$ maximal and $\theta_{13}$ small at the same time. 
The Renormalization Group (RG) evolution is a 
possible candidate for accomplishing this.
Therefore, it is interesting to investigate the evolution of the mixing angles
from the GUT scale to the electroweak (EW) or SUSY-breaking scale.
A number of studies with three neutrinos considered the
possibility of increasing a small mixing angle via RG evolution
\cite{Tanimoto:1995bf,Balaji:2000au,Miura:2000bj,Dutta:2002nq}.
Others focused on the case of nearly degenerate neutrinos
\cite{Ellis:1999my,Casas:1999tp,Casas:1999ac,Chankowski:2000fp,Chen:2001gk,Parida:2002gz},
on the existence of fixed points \cite{Chankowski:1999xc}, 
or on the effect of non-zero Majorana phases on
the stability of the RG evolution \cite{Haba:2000tx}.

We consider the see-saw scenario, i.e.\ the Standard Model (SM) or MSSM
extended by 3 heavy neutrinos that are singlets under the SM gauge
groups and have large explicit (Majorana) masses with a non-degenerate
spectrum.
Due to this non-degeneracy, one has to use several effective
theories, with the singlets partly integrated out, when studying the
evolution of the effective mass matrix of the light neutrinos
\cite{King:2000hk,Antusch:2002rr}.
Below the lowest mass threshold, the neutrino mass matrix is given
by the effective dimension 5 neutrino mass operator in the SM or MSSM. 
The relevant RGE's were derived in 
\cite{Antusch:2002rr,Chankowski:1993tx,Babu:1993qv,Antusch:2001ck,Antusch:2002vn,Antusch:2002ek}.   

In this paper, we assume bimaximal mixing at the GUT scale with
vanishing CP phases and positive mass eigenvalues.
We calculate the RG
running numerically in order to obtain the mixing angles at low energy
and to compare them with the experimentally favored values.
We include the regions above and between the see-saw scales in our study,
which have not been considered in most of the previous works.
We find that the solar mixing angle changes considerably, while the
evolution of the other angles is comparatively small,
so that values compatible with the LMA solution can be obtained.  We
present analytic approximations that help to understand this behavior
and show that it is rather generic.

\vspace*{-0.5cm}
\section{Bimaximal Mixing at the GUT Scale}
\begin{figure}[h]
 \begin{center}
  \begin{subfigure}[Normal hierarchy]{\label{subfig:NeutrinoMassScheme1}
   \CenterEps{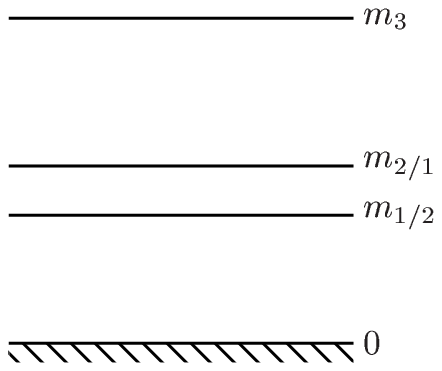}}
  \end{subfigure}
  \hfil
  \begin{subfigure}[Inverted hierarchy]{\label{subfig:NeutrinoMassScheme2}
  	\CenterEps{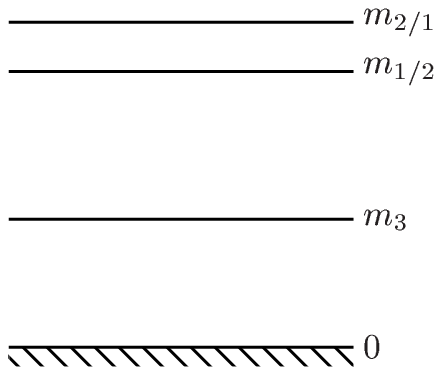}}
  \end{subfigure}
 \end{center}
 \caption{Possible mass hierarchies for the light neutrinos.  
 We use the convention that $m_1$ and $m_2$ are chosen in such a way
 that $0 \le \theta_{12} \le 45^\circ$.
 The LMA solution then requires $m_2 > m_1$.}
 \label{fig:NeutrinoMassSchemes}
\end{figure}

At the GUT scale, we assume bimaximal mixing in the lepton sector.
We restrict ourselves to the case of positive mass eigenvalues and real
parameters, so that there is no CP violation. 
In the basis where the charged lepton Yukawa matrix
is diagonal, up to phase conventions the general parametrization of
the effective Majorana mass matrix of the light neutrinos is then 
\begin{eqnarray} \label{eq:BimaxFormofMnu}
 m^{\mathrm{bimax}}_\nu 
 = 
 V(\tfrac{\pi}{4}, 0, \tfrac{\pi}{4}) \cdot \diag(m_1,m_2,m_3)
 \cdot V^T(\tfrac{\pi}{4}, 0, \tfrac{\pi}{4})
 =
 \left(\begin{array}{ccc}
  a\!-\!b\: & c & -c\\
  c & a & b\\
  -c & b & a\\
 \end{array}\right)
\end{eqnarray}
where
\begin{equation}
 V(\theta_{12},\theta_{13},\theta_{23})=\left(
 \begin{array}{ccc}
 c_{12}c_{13} & s_{12}c_{13} & s_{13}\\
 -c_{23}s_{12}-s_{23}s_{13}c_{12} &
 c_{23}c_{12}-s_{23}s_{13}s_{12} & s_{23}c_{13}\\
 s_{23}s_{12}-c_{23}s_{13}c_{12}&
 -s_{23}c_{12}-c_{23}s_{13}s_{12} & c_{23}c_{13}
 \end{array}
 \right)
\end{equation}
with \(s_{ij}=\sin\theta_{ij}\) and \(c_{ij}=\cos\theta_{ij}\)
is the (orthogonal) CKM matrix in standard parametrization, and
\begin{subequations}\label{eq:abcAsFunctionsOfm1m2m3}
\begin{eqnarray}
 a & = & \frac{1}{4}\left(m_1+m_2+2\,m_3\right) \;,
 \label{eq:aAsFunctionOfm1m2m3}
 \\
 b & = & \frac{1}{4}\left(-m_1-m_2+2\,m_3\right) \;,
 \\
 c & = & \frac{m_2-m_1}{2\sqrt{2}} \;.
\end{eqnarray}
\end{subequations}
Inverting equations \eqref{eq:abcAsFunctionsOfm1m2m3} yields the mass
eigenvalues
\begin{subequations}\label{eq:m1m2m3AsFunctionsOfabc}
\begin{eqnarray}
 m_1 &=& a-b - \sqrt{2}\,c \;,
 \\*
 m_2 &=& a-b + \sqrt{2}\,c \;,
 \\*
 m_3 &=& a+b \;.
\end{eqnarray}
\end{subequations}
From equation (\ref{eq:aAsFunctionOfm1m2m3}) we see that $a>0$.
Equations (\ref{eq:m1m2m3AsFunctionsOfabc}) imply that the 
solar mass squared difference \(\Delta m_\mathrm{sol}^2=m_2^2-m_1^2\)
is related to $c$, while the atmospheric one,
\(\Delta m_\mathrm{atm}^2=m_3^2-m_1^2\), is controlled by $b$. 
Thus, $a>|b|>|c|$. 
For $b>0$ we obtain a normal mass hierarchy, 
while for $b<0$ the mass hierarchy is inverted, as 
illustrated in figure \ref{fig:NeutrinoMassSchemes}.
For positive $c$, 
\(m_1<m_2\), otherwise \(m_1>m_2\). Hence,
\(\Delta m_\mathrm{sol}^2\) is positive only if \(c\) is.
If $a \gg |b|,|c|$, the spectrum is called degenerate.
We use the convention that the mass label 2 is attached in such a way
that $0 \leq \theta_{12} \leq 45^\circ$.
This can always be accomplished by the replacement 
\(c \leftrightarrow -c\).

In our see-saw scenario, the effective mass matrix of the light
neutrinos is
\begin{equation}\label{eq:SeeSawF}
	m_\nu^\mathrm{bimax} =
	\frac{v_\mathrm{EW}^2}{2} \, Y_\nu^T M^{-1} Y_\nu 
\end{equation}
at the high-energy scale,
with $\<\phi\> = \frac{v_\mathrm{EW}}{\sqrt{2}} \approx 174$ GeV.
Obviously, the singlet Yukawa and mass matrices $Y_\nu$ and $M$ cannot
be determined uniquely from this relation, i.e.\ there is a set of
$\{Y_\nu, M\}$ configurations that yield bimaximal mixing.
After choosing an initial condition for $Y_\nu$,
$M$ (and thus the see-saw scales) is fixed by 
the see-saw formula (\ref{eq:SeeSawF}) if $Y_\nu$ is invertible.

\vspace*{-0.5cm}
\section{Solving the RGE's} 
To study the RG running of the leptonic mixing angles and neutrino
masses, all parameters of the
theory have to be evolved from the GUT scale to the EW or 
SUSY-breaking scale, respectively. Since the heavy singlets have to be integrated out
at their mass thresholds, 
which are non-degenerate in general,
a series of effective theories has to be used.
The derivation of the RGE's
and the method for dealing with these effective theories 
are given in \cite{Antusch:2002rr}. Starting at
the GUT scale, the strategy is to successively solve the systems
of coupled differential equations of the form
\begin{eqnarray}
16\pi^2 \, \mu \frac{\D}{\D \mu}   \accentset{(n)}{X}_i
  = \accentset{(n)}{\beta}_{{X}_i} \RaiseBrace{\bigl(}\RaiseBrace{\bigl\{}
  \accentset{(n)}{X}_j\RaiseBrace{\bigl\}}\RaiseBrace{\bigl)}
\end{eqnarray}
for all the parameters $\accentset{(n)}{X}_i, \accentset{(n)}{X}_j \in 
\RaiseBrace{\bigl\{}\accentset{(n)}{\kappa},\accentset{(n)}{Y_\nu},\accentset{(n)}{M},
\dots\RaiseBrace{\bigr\}}$  
of the theory
in the energy ranges corresponding to the effective theories denoted by
$(n)$. At each see-saw scale, tree-level matching 
is performed.
Due to the complicated structure of the set of differential equations, 
the exact solution can only be obtained numerically.
However, to understand certain features of the RG evolution,
an analytic approximation at the GUT scale will be derived in section
\ref{sec:AnalyticApproximation}.

\vspace*{-0.5cm}
\section{Examples for the Running of the Mixing Angles}
\begin{figure}[h]
\vspace*{2mm}
        \begin{center}              
		\CenterEps{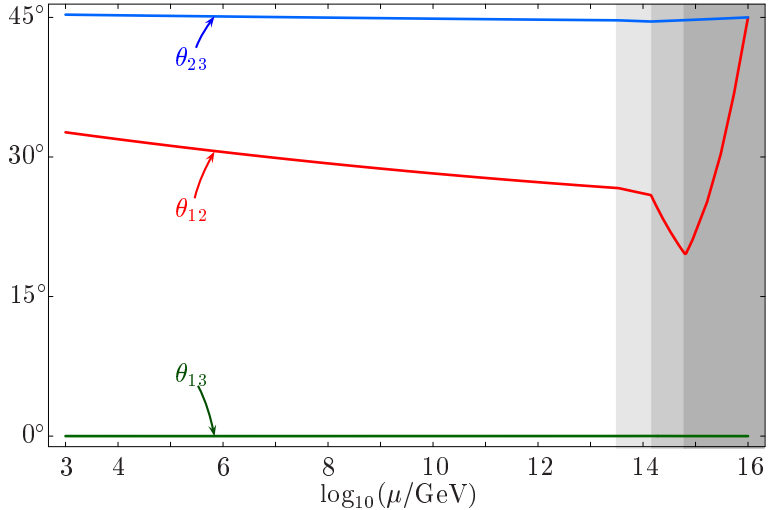}
        \end{center}
\vspace*{2mm}
\caption{\label{fig:3} 
RG evolution of the mixing angles
from the GUT scale to the SUSY-breaking scale (taken to be $\approx 1$~TeV) 
in the MSSM extended by heavy singlets for a normal mass hierarchy and 
$Y_\nu = X \, \diag\left( 1, \varepsilon, \varepsilon^2 \right)$
with $\tan\beta = 5$, $\varepsilon=0.525$, $a=0.0675$~eV and $X=1$.
In this example, the lightest neutrino has a mass of $0.025$~eV. 
The kinks in the plots 
correspond to the mass thresholds at the see-saw scales. 
The grey-shaded regions mark the various effective theories.
}
\end{figure}

\begin{figure}[h]
       \begin{center}                   
		\CenterEps{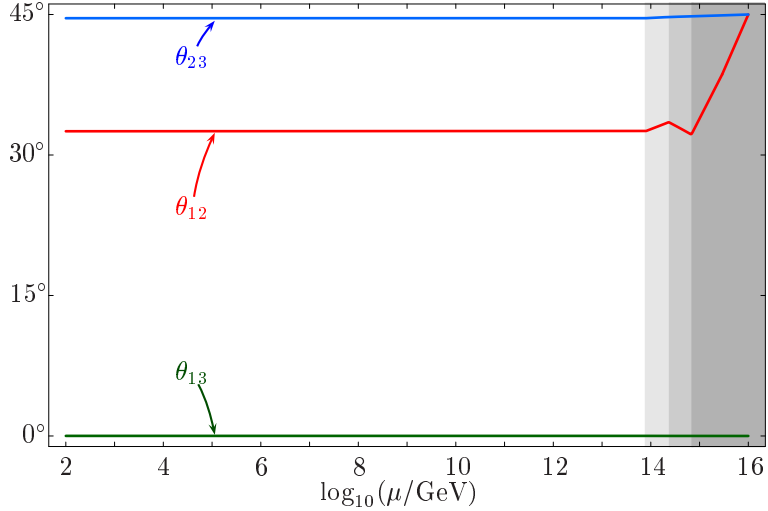}
       \end{center}
\vspace*{2mm}
\caption{\label{fig:2} 
Example for the RG evolution of the mixing angles 
in the SM extended by heavy singlets 
from the GUT scale to the EW scale for a normal mass hierarchy and
$Y_\nu = X \, \diag\left( 1, \varepsilon, \varepsilon^2 \right)$
with $\varepsilon=0.65$, $a=0.0655$~eV and $X=1$. 
In this example, the lightest neutrino has a mass of $0.024$~eV.  
}        
\end{figure}

Figures \ref{fig:3} and \ref{fig:2} show typical numerical 
examples for the
running of the mixing angles from the GUT scale to the EW or
SUSY-breaking scale. 
They contain an important effect that appears for most choices of the
initial parameters:
The solar angle $\theta_{12}$ changes drastically, while the changes in 
$\theta_{13}$ and $\theta_{23}$ are comparatively small.
This agrees remarkably well with the experimentally favored scenario.


\vspace*{-0.5cm}
\section{Analytic Approximation for the Running of the Mixing Angles at the GUT
Scale} \label{sec:AnalyticApproximation}\label{sec:AnalyticApprox}
In order to understand the effect found numerically in the previous
section, we now derive an analytic approximation for the RG evolution of
the mixing angles at the GUT scale.
It is only affected by the part of the RGE that is not proportional to
the unit matrix, which is given by
 \begin{eqnarray} \label{eq:RGEforGamma}
  16\pi^2 \, \mu \frac{\D}{\D \mu} m_\nu
  & = & 
  C_e \left[Y_e^\dagger Y_e\right]^T m_\nu +
  C_e\,m_\nu \left[Y_e^\dagger Y_e\right] 
 \nonumber \\*
  &&{}+ 
  C_\nu \left[Y_\nu^\dagger Y_\nu\right]^T m_\nu +
  C_\nu\,m_\nu \left[Y_\nu^\dagger Y_\nu\right]
 \nonumber \\*
  &&{}+ \;\mbox{terms with trivial flavour structure}
  \label{eq:RunningOfCompositeMassOperator}
 \end{eqnarray}
 with $C_e = -\tfrac{3}{2}$, $C_\nu = \tfrac{1}{2}$ in the SM and
 $C_e = C_\nu = 1$ in the MSSM.
Analogously to equation \eqref{eq:BimaxFormofMnu}, $m_\nu$ is
parametrized by
 \begin{equation}\label{eq:mParamatrization}
  m_\nu(t)
  = 
  V\big(\theta_{12}(t),\theta_{13}(t),\theta_{23}(t)\big)
  \cdot m_\mathrm{diag}(t)\cdot 
  V^T\big(\theta_{12}(t),\theta_{13}(t),\theta_{23}(t)\big) \;,
 \end{equation}
where $\mu$ is the renormalization scale, $t := \ln \frac{\mu}{\mu_0}$,
and $m_\mathrm{diag} := \diag(m_1,m_2,m_3)$.
In general, the real $Y_\nu$ can be written as
\begin{equation} \label{eq:YNuParam}
  Y_\nu=
  V(\xi_{12},\xi_{13},\xi_{32})
  \cdot \diag(y_1,y_2,y_3)  \cdot
  V^T(\phi_{12},\phi_{13},\phi_{32}) \;.
\end{equation}
However, the effective mixing matrix $m_\nu$ is 
invariant under the transformations
$Y_\nu \rightarrow V^T Y_\nu$ and $M \rightarrow V^T M V$, which correspond to a
change of basis for the heavy sterile neutrinos.
Thus, $V(\xi_{12},\xi_{13},\xi_{32})$
in equation (\ref{eq:YNuParam}) can be absorbed into $M$, 
leading to the simpler parametrization 
 \begin{equation}\label{eq:GeneralParametrizationOfYnu}
  Y_\nu(y_1,y_2,y_3,\phi_{12},\phi_{13},\phi_{32}) 
  =
  \diag(y_1,y_2,y_3)  \cdot
  V^T(\phi_{12},\phi_{13},\phi_{32})\;.
 \end{equation}

Furthermore, we use the approximation that the effect of the charged
lepton Yukawa matrices $Y_e$ can be neglected compared to that of the
neutrino Yukawa matrix. 
Note that in the MSSM a large $\tan \beta$ can yield a relatively large
$Y_e$, which can also have sizable effects that are neglected in this
approximation.

We now differentiate equation \eqref{eq:mParamatrization} w.r.t.\ \(t\)
and insert the RGE \eqref{eq:RunningOfCompositeMassOperator}.
For the evolution of the mixing angles at the GUT scale with bimaximal
mixing as initial condition, we thus obtain both in the SM and in the
MSSM the ratios
 \begin{subequations} \label{eq:AnalyticApprox}
 \begin{eqnarray}
  \left. \frac{\Dot{\theta}_{12}}{\Dot{\theta}_{13}} \right|_{M_\mathrm{GUT}}
  & = & 
  \frac{2\,{\sqrt{2}}\,\left( m_{1} + m_{2} \right) \,
     \left( m_{3} - m_{1} \right) \,
     \left( m_{3} - m_{2} \right) \, \,
  F_{1}
  }{\left( m_{2} - m_{1} \right) \,\left[ 8\,\left( m_{3}^2-m_{1}\,m_{2} \right) \,
   F_{2}
   +  4\,{\sqrt{2}}\,\left( m_{2} - m_{1} \right) \, m_{3}\,
   F_{3}
   \right] } 
 \nonumber \\*[2mm]
  & \approx &
  \left\{\begin{array}{ll}
   \displaystyle \pm \frac{1}{2\sqrt{2}} \,
   \frac{m_2+m_1}{m_2-m_1} \,\frac{F_1}{F_2}&
   \mbox{for hierarchical neutrino masses\footnotemark[1]}
  \\[3mm]
   \displaystyle \frac{1}{2\sqrt{2}} \,
   \frac{\Delta m_\mathrm{atm}^2}{\Delta m_\mathrm{sol}^2} \,\frac{F_1}{F_2}&
   \mbox{for degenerate neutrino masses}
  \end{array}\right. 
 \\[3mm] 
  \left. \frac{\Dot{\theta}_{12}}{\Dot{\theta}_{23}} \right|_{M_\mathrm{GUT}}
  & = & 
  \frac{2\,{\sqrt{2}}\,\left( m_{1} + m_{2} \right) \,
    \left( m_{3} - m_{1} \right) \,
     \left( m_{3} - m_{2} \right) \,
 	F_{1}}{\left( m_{2} - m_{1} \right) \,
    \left[ 8\,\left( m_{2} - m_{1} \right) \,
       m_{3}\,  F_{2}
  	+      4\,{\sqrt{2}}\,\left( m_{3}^2 - m_{1}\,m_{2}\right)
  F_{3}\right] }
 \nonumber \\*[2mm]
  & \approx &
  \left\{\begin{array}{ll}
  \displaystyle \pm \frac{1}{2} \,\frac{m_2+m_1}{m_2-m_1} \,\frac{F_1}{F_3}&
  \mbox{for hierarchical neutrino masses\footnotemark[1]}
 \\[3mm]
  \displaystyle \frac{1}{2} \,\frac{\Delta m_\mathrm{atm}^2}{\Delta m_\mathrm{sol}^2} \,\frac{F_1}{F_3}&
  \mbox{for degenerate neutrino masses}
  \end{array}\right.
 \end{eqnarray}
 \end{subequations}
   \footnotetext[1]{
	Note that this approximation is also valid for a relatively weak
	hierarchy, where $m_3$ is a few times larger or smaller than $m_1$, 
	$m_2$.
   }
 with
 \begin{subequations}
 \begin{eqnarray}
  F_1
  &=& 
  \left( y_{1}^2 - y_{2}^2 \right) \,\left\{\vphantom{y_1^2}
       \cos (2\,\phi_{12})\,
  	\left[\vphantom{y_1^2}
    	\left(  \cos (2\,\phi_{13}) -3\right) \,\sin (2\,\phi_{23}) 
		-6\,\cos^2 (\phi_{13}) \right]\right.
   \nonumber\\*
   && {}
  \left.\hphantom{\left( y_{1}^2 - y_{2}^2 \right) \,\left\{\right.}
  -4\, \cos (2\,\phi_{23})\,\sin (2\,\phi_{12})\,
       \sin (\phi_{13}) \vphantom{y_1^2}\right\}
  \nonumber\\*
  & &
  +      \left( y_{1}^2 + y_{2}^2 
  -         2\,y_{3}^2 \right) \,\left[\vphantom{y_1^2}\cos (2\,\phi_{13})\,
       \left(  \sin (2\,\phi_{23}) -3\right)  
  +       \left( 1 + \sin (2\,\phi_{23}) \right)  \vphantom{y_1^2}\right]
 \,,
 \nonumber\\*
 & & \\
  F_2
  &= &
  2\,\left( y_{1}^2 - y_{2}^2 \right) \,
          \cos (\phi_{13})\,\sin (2\,\phi_{12})\,
          \left( \sin (\phi_{23}) -\cos (\phi_{23}) \right) 
  \nonumber\\*
  & & 
  -         \left( y_{1}^2 + y_{2}^2 
  -            2\,{y_{3}}^2 
  +            \left( y_{1}^2 - y_{2}^2 \right) \,
             \cos (2\,\phi_{12}) \right) \times\nonumber\\*
  & &{}\qquad\times
          \sin (2\,\phi_{13})\,
          \left( \cos (\phi_{23}) 
  +            \sin (\phi_{23}) \right) 
  \;,\\
  F_3
  & = &
  \left( y_{1}^2 - y_{2}^2 \right)  \,\left[\vphantom{y_1^2}
             \cos (2\,\phi_{12})\,
             \left( \cos (2\,\phi_{13})-3 \right)   \,
          \cos (2\,\phi_{23})\right.
  \nonumber\\*
  & & 
  \hphantom{  \left( y_{1}^2 - y_{2}^2 \right)  \,\left[\right.}\left.
  {}+         4\, \sin (2\,\phi_{12})\,\sin (\phi_{13})\,
          \sin (2\,\phi_{23})\vphantom{y_1^2}\right]
  \nonumber\\*
  & &{}
  + 2\,\left( y_{1}^2  +  y_{2}^2 - 2\,y_{3}^2 \right) \,
             \cos^2 (\phi_{13})\, \cos (2\,\phi_{23}) \;.
 \end{eqnarray}
 \end{subequations}
This result can also be obtained from the formulae derived in
\cite{Casas:1999tg}.
The constants $F_1$, $F_2$ and $F_3$ clearly depend on the choice of
$Y_\nu (M_\mathrm{GUT})$.  However, unless the parameters 
$\{y_1,y_2,y_3,\phi_{12},\phi_{13},\phi_{32}\}$ are fine-tuned, we
expect the ratios $F_1 / F_2$ and $F_1 / F_3$ to be of the order one. 
Consequently, the RG change of $\theta_{12}$ is larger than that of the
other angles if the mass-dependent factors in equations
\eqref{eq:AnalyticApprox} are large.  This is always the case for
degenerate neutrino masses, since 
$\Delta m^2_\mathrm{atm} \gg \Delta m^2_\mathrm{sol}$.
As \((m_1\!-\!m_2)\) is related to the small solar mass squared difference,
it is also true for non-degenerate mass schemes, unless $m_1$ is very
small, in which case the ratio approaches 1.  This corresponds to
a normal mass hierarchy and a strongly hierarchical mass scheme.  
Finally, it can be shown that
the running of \(\theta_{12}\) is always enhanced compared to
that of \(\theta_{13}\) and \(\theta_{23}\) for inverted schemes.
Hence, we conclude that this is a generic effect.

\vspace*{-0.5cm}
\section{Parameter Space Regions Compatible with the LMA Solution}
\label{sec:NumResults}
\vspace*{-2mm}
\subsection{Parameters at the GUT Scale}
\begin{figure}[h]
\begin{center}
\subfigure[Parameter $a$
]{\label{subfig:Param-a}
\(\CenterEps[1]{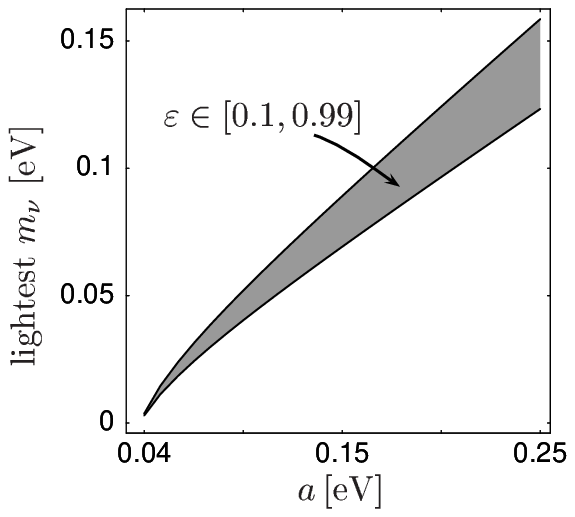}\)
}
\hfil
\subfigure[Parameter $\varepsilon$
]{\label{subfig:Param-e}
\(\CenterEps[1]{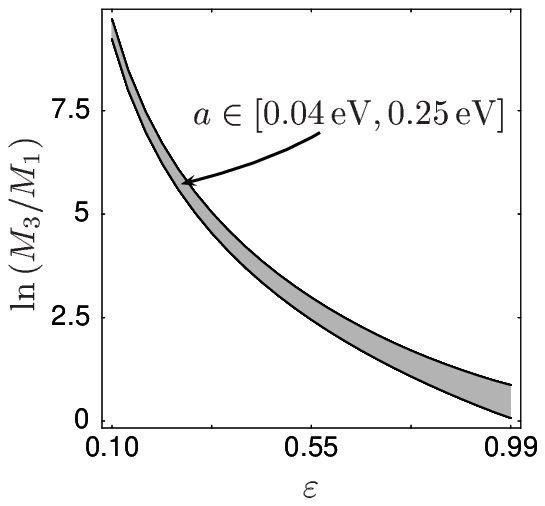}\)
}
\end{center}
\caption{
 Plot \ref{subfig:Param-a} shows the
 mass of the lightest neutrino (at low energy)
 as a function of $a$ for the SM and the MSSM with normal mass
 hierarchy, $X=1$ and $\varepsilon \in [0.1,0.99]$ (grey region). 
 Plot \ref{subfig:Param-e} shows the degeneracy of the see-saw scales,
 parametrized by $\ln (M_3/M_1)$ (at the GUT scale), as a
 function of $\varepsilon$ for the same cases with
 $a \in [0.04\text{ eV}, 0.25\text{ eV}]$ (grey region).
}
\label{fig:Parameters}
\end{figure}

\vspace*{-2mm}
The considerable change of the solar mixing angle found in the previous
sections 
raises the question whether the parameter region of the LMA
solution might be reached by RG evolution, if one starts with bimaximal
mixing at high energy.  We will investigate this possibility by further
numerical calculations in the following.
To reduce the parameter space for the numerical analysis, we
choose a specific neutrino Yukawa coupling $Y_\nu$ at the GUT scale.
We assume that it is diagonal and of the form
\begin{equation} \label{eq:YnuDiagforBimax}
	Y_\nu = X \, \diag\left( 1, \varepsilon, \varepsilon^2 \right) \;.
\end{equation}
$Y_\nu$ and $M$ are now determined by the parameters $\{\varepsilon,X,
a,b,c\}$.
Moreover, we fix the GUT scale values of $b$ and $c$ by the requirement
that the solar and atmospheric mass squared differences obtained at the
EW scale after the RG evolution be compatible with the allowed
experimental regions.  Thus, we are left with the free parameters $X$,
$\varepsilon$ and $a$.
The parameter $\varepsilon$ controls the hierarchy of the entries in
$Y_\nu$ and thus the degeneracy of the see-saw scales,
while $a$ determines the mass of the lightest neutrino.
The dependence of physical quantities on $\varepsilon$ and $a$ 
is shown in figure \ref{fig:Parameters}.
The effect of changing the scale $X$ 
of the neutrino Yukawa coupling will be
discussed in section \ref{sec:SmallX}.
As mentioned above, we work in the basis where the Yukawa matrix of
the charged leptons is diagonal.

\subsection{Allowed Parameter Space Regions}
\begin{figure}[h]
\begin{center}
\subfigure[SM
]{\label{subfig:SMr1Yinv}\(\CenterEps[1]{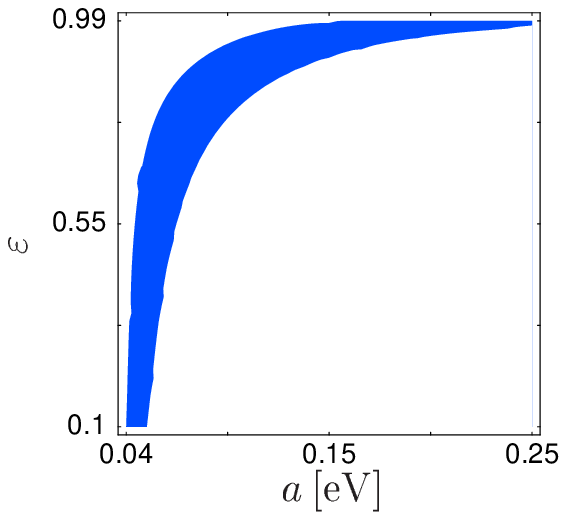}\)}
\hfil
\subfigure[MSSM
]{\label{subfig:MSSMr1Yinv}\(\CenterEps[1]{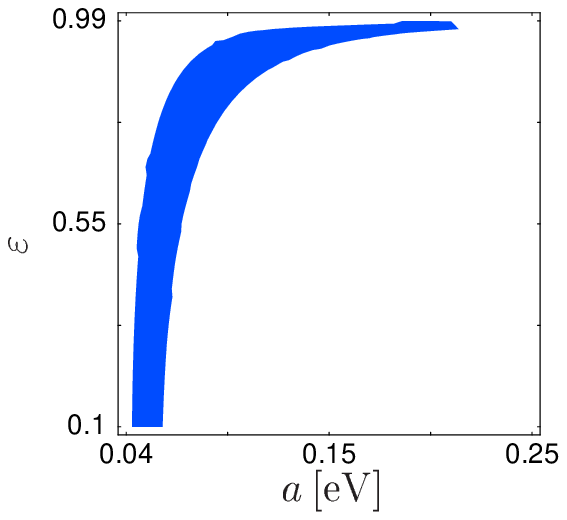}\)}
\end{center}
\vspace{-1mm}
\caption{\label{fig:PScans} 
Parameter space regions compatible with the LMA solution of the solar neutrino
problem for the example $Y_\nu = \diag(1, \varepsilon, \varepsilon^2)$.
The initial condition at the GUT scale $M_{\mathrm{GUT}}=10^{16}$ GeV is
bimaximal mixing, and the comparison with the experimental data is performed at
the EW scale or at 1~TeV for the SM 
and the MSSM, 
respectively.
The white regions of the plots are excluded by the data (LMA) at $3 \sigma$.
For this example, we consider the case of a normal neutrino mass hierarchy and
$X=1$ for the scale factor of the neutrino Yukawa couplings.
}
\vspace*{6mm}
\end{figure}

The parameter space regions 
in which the RG evolution produces low-energy values
compatible with the LMA solution 
are shown in figure \ref{fig:PScans} for the SM and the MSSM ($\tan
\beta = 5$) with a normal mass hierarchy. 
We find that for the form of $Y_\nu$ under consideration, 
hierarchical and degenerate neutrino mass schemes as well as degenerate
and non-degenerate see-saw scales are possible.    
For inverted neutrino mass spectra, allowed
parameter space regions exist as well. 

We would like to stress that the shape of
the allowed parameter space regions strongly depends on the choice of 
the initial value of $Y_\nu$ at the GUT scale. One also has to ensure that the
sign of $\Delta m^2_{\mathrm{sol}}$ is positive, as the LMA solution
requires this if the convention is used that the solar mixing angle is
smaller than $45^\circ$. 
With bimaximal mixing at the GUT scale, the sign of 
$\Delta m^2_{\mathrm{sol}}$ is not defined by the initial conditions.
Using the analytic approximation of section
\ref{sec:AnalyticApproximation}, the sign 
just below the GUT scale can be calculated. 
We find $\Delta m^2_{\mathrm{sol}} > 0$ for $F_1 <0$ and vice versa. 
However, in order to  
predict the sign of $\Delta m^2_\mathrm{sol}$ at low energy,
the numerical RG evolution has to be used.
This excludes some of the possible choices for the neutrino
Yukawa coupling $Y_\nu$ at the GUT scale. For example, among the possibilities
with diagonal $Y_\nu$ it excludes
$Y_\nu = \diag(\varepsilon^2, \varepsilon, 1)$.

\subsection{Dependence on the Scale $X$ of the Neutrino Yukawa Coupling}
\label{sec:SmallX}
\begin{figure}[h]
       \begin{center}
		\CenterEps{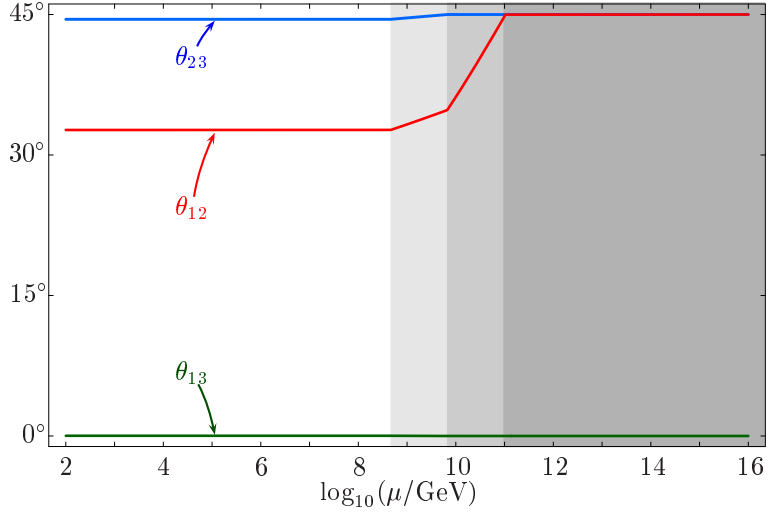}
       \end{center}
\caption{\label{fig:RunningForSmallX} 
RG evolution in the SM for $X=0.01$, $\varepsilon=0.3$, $a=0.0535$~eV 
and a normal mass hierarchy. 
The running from bimaximal mixing to the LMA solution 
now takes place between the see-saw scales.  
In this example, the lightest neutrino has a mass of $0.017$~eV. 
}        
\end{figure}

\vspace{-2mm}
For small values of $X$, the contribution from $Y_\nu$ to the 
evolution of the mixing angles above the largest see-saw scale 
is suppressed by a factor of $X^2$. 
Nevertheless, the evolution 
to the LMA solution is still possible,
as can be seen from the example in figure \ref{fig:RunningForSmallX}. 
Here the large change of $\theta_{12}$ also seems to be generic but
takes place between the
see-saw scales, which shows the importance of carefully studying the RG
behavior in these intermediate regions \cite{Antusch:2002rr}. 
Note that in this case the analytic approximation of section
\ref{sec:AnalyticApprox} cannot be applied, since it is only valid at
the GUT scale.

\subsection{Effect of Neutrino CP Parities}
\begin{figure}[h]
       \begin{center}
		\CenterEps{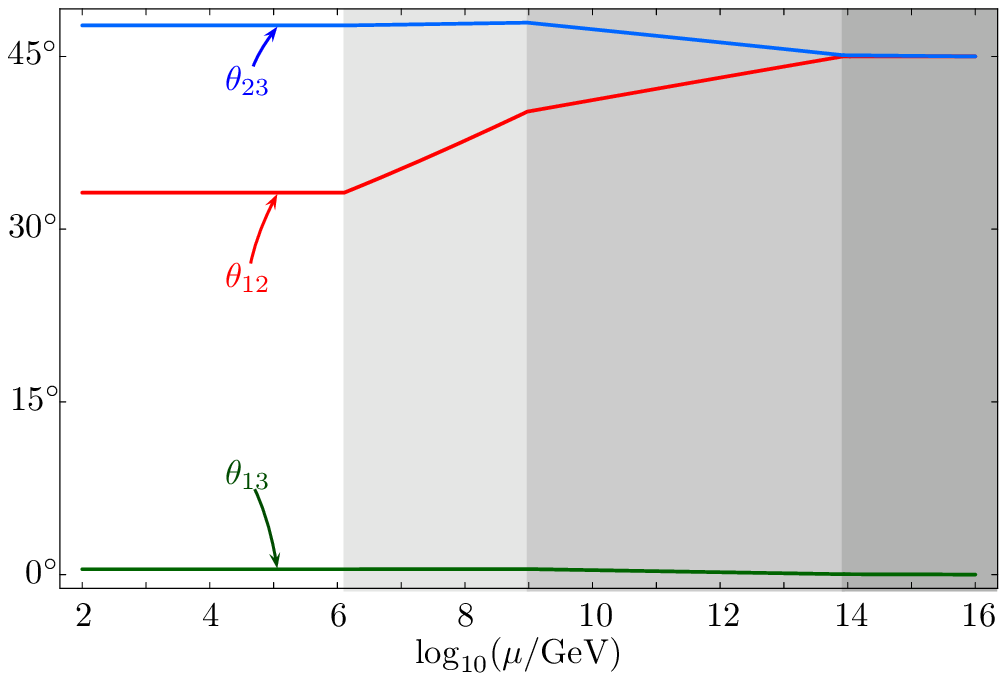}
       \end{center}
\caption{\label{fig:RunningWithMajParities} 
RG evolution in the SM with a negative CP parity
for $m_2$, $X=0.5$, $\varepsilon=3.5 \cdot 10^{-3}$ and 
a normal mass hierarchy. 
The running from bimaximal mixing to the LMA solution 
takes place between the see-saw scales.  
In this example, we consider a strongly hierarchical mass spectrum. 
The lightest neutrino has a mass of $0.004$~eV. 
Note that the cases $\theta_{23} > 45^\circ$, $\Delta m^2_{23} > 0$ and
$\widetilde{\theta}_{23} := 90^\circ - \theta_{23} < 45^\circ$, 
$\Delta \widetilde{m}^2_{23} := -\Delta m^2_{23}$ are indistinguishable
in neutrino oscillations.
}        
\end{figure}

\vspace{-2mm}
An example for the
running of the mixing angles to the LMA solution with a negative CP parity
for the state with mass $m_2$ is shown in figure \ref{fig:RunningWithMajParities}. For this
we have chosen a different diagonal structure for $Y_\nu$, 
\begin{equation} \label{eq:YnuDiag2forBimax}
	Y_\nu = X \, \diag\left( \varepsilon^2 , \varepsilon, 1 \right) \;,
\end{equation}
at the GUT scale.  Here, the evolution to the LMA solution is possible
due to running between the see-saw scales.  A more detailed study of the effect of CP phases    
will be given in a forthcoming paper \cite{Antusch:2002pr}.

The large RG effects in this case seem surprising at first sight, since
previous studies, e.g.\ \cite{Haba:2000tx,Casas:1999tg}, found that
opposite CP parities for $m_1$ and $m_2$ prevent a sizable change
of the solar mixing angle by RG evolution.  However, these works did not
consider the energy region between the see-saw scales, where the largest
change occurs in our example.  This fact explains the apparent
discrepancy.

\subsection{Low Scale Values of $\theta_{13}$ and $\theta_{23}$}
\vspace{-2mm}
The mixing angles $\theta_{13}$ and $\theta_{23}$ are affected by the RG
evolution as well, i.e.\ they do not stay at their initial values 
$\theta_{13}=0^\circ$ and $\theta_{23}=45^\circ$.
However, lower bounds on their changes cannot be given
unless a specific model is chosen.  As one can see from the previous
examples,
the changes
can be tiny. For instance, the evolution of figure \ref{fig:3} gives
$\Delta \theta_{13} = 0.02^\circ$, which corresponds to $\sin^2 (2 \theta_{13})=5
\cdot 10^{-7}$, and $\Delta\theta_{23} = 0.28^\circ$. On the other hand,
other choices of $Y_\nu$
at the GUT scale produce $\Delta \theta_{13}$ and $\Delta \theta_{23}$
that come close to the experimental bounds. This can make it possible to
discriminate between models with different initial values 
$Y_\nu(M_\mathrm{GUT})$.

\section{Summary and Conclusions}
\vspace{-2mm}
We have shown that in see-saw scenarios the experimentally
favored neutrino mass parameters with the LMA solution of the solar
neutrino problem can be obtained in a rather generic way from bimaximal
mixing at the GUT scale by Renormalization Group running. 
We have concentrated on the case of vanishing CP phases, which implies
positive mass eigenvalues.
In an example where the mass eigenstates corresponding to the solar mass
squared difference have opposite CP parity, we
have demonstrated that an evolution towards the LMA solution is possible
in this case as well.
The general case of arbitrary CP phases is beyond the scope of this
letter and will be studied elsewhere \cite{Antusch:2002pr}.
The mixing angles evolved down to the electroweak scale show a strong
dependence on the mass scale of the lightest neutrino, on the degeneracy
of the see-saw scales, and on the form of the neutrino Yukawa coupling. 
A generic feature of the Renormalization Group evolution is that the
solar mixing angle $\theta_{12}$ evolves towards sizably smaller values,
whereas the change of $\theta_{13}$ and $\theta_{23}$ is comparatively small.
In the SM and MSSM, we find extensive regions in parameter
space which are compatible with the LMA solution for normal and inverted
neutrino mass hierarchies and for large and small absolute scales 
of the neutrino Yukawa couplings.  
Thus, RG running may provide a natural explanation for the observed 
deviation of the LMA mixing angles from bimaximality.

\vspace{-1mm}
\ack
\vspace{-2mm}
We would like to thank P.~Huber for useful discussions. 
This work was supported in part by the 
``Sonderforschungsbereich~375 f\"ur Astro-Teilchenphysik der 
Deutschen Forschungsgemeinschaft''.

\end{document}